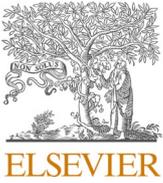
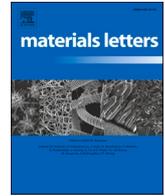

# Stress-induced phase transformations in Ti-15Mo alloy at elevated temperature

Petr Harcuba [a], Tomáš Krajňák [a,b,*], Dalibor Preisler [a], Jiří Kozlík [a], Josef Stráský [a], Jana Šmilauerová [a], Fernando Gustavo Warchomicka [c], Miloš Janeček [a]

[a] *Charles University, Faculty of Mathematics and Physics, Ke Karlovu 3, 12116 Prague, Czech Republic*
[b] *Research Centre, University of Žilina, Univerzitná 1, 01026 Žilina, Slovakia*
[c] *Graz University of Technology, Inst Mat Sci Joining & Forming, A-8010 Graz, Austria*

A B S T R A C T

Controlled mechanical loading was applied to Ti-15Mo alloy during annealing at 550 °C. Massive formation of the $\omega_{iso}$ phase from the parent β-phase occurred during annealing without external stress or with stress well below the yield stress. Moreover, a massive α phase precipitation takes place under simultaneous annealing and plastic deformation. Plastic deformation plays a key role in β → α transformation and achieving refined α + β type microstructure resulted in improved mechanical properties. Studying phase transformations during plastic deformation is critical for understanding and optimizing thermomechanical processing of metastable β-Ti alloys.

## 1. Introduction

Metastable β-Ti alloys undergo multiple phase transformations during annealing below the β-transus temperature or upon cooling to room temperature [1,2]. ω phase growth [3] and α phase precipitation in the β matrix [4] are typical diffusion-driven transformations at elevated temperatures. Metastable β-Ti alloys also experience stress-induced and deformation-induced transformations, namely formation of stress-induced martensite.

The deformation below the β-transus and associated phase transformations with initial single β phase microstructure have been reported only rarely. Thermo-mechanical processing of Ti-6246 alloy cooled from the β field showed deformation induced α formation, affected by defects acting as nucleation sites [5,6]. Deformation induced α phase was also observed with beta annealed microstructure heated up to 600–700 °C in Ti-7333 [7,8] and Ti-5553 [9,10] with very short soaking time. The authors observed a dynamic precipitation of α, following selected α variant to nucleate in the beta field.

Among the β-Ti alloys, Ti-15Mo represents a simple binary system that has been extensively studied in the last decade. Numerous studies focused on diffusional phase transformations during/after annealing without external stress [11,12]. Yao et al. studied phase transformation under deformation, observing the decreasing twinning upon temperature increase (up to 400 °C) [13]. Wang et al. investigated stress-induced β → ω phase transformations in cold-drawn Ti-15Mo wires [14].

This study examines stress- and deformation-induced phase transformations in Ti-15Mo at 550 °C. The temperature was selected as it is just below the ω-solvus temperature, while a complete dissolution of $\omega_{iso}$ particles occurs at the temperature of 560 °C [15,16]. In a stress/strain-free condition, α phase does not precipitate upon the annealing treatment used in this study, allowing us to reveal the effect of stress and deformation on the β → α transformation.

## 2. Experimental procedure

Ti-15Mo alloy (Carpenter Technology Co., USA) contained 15 wt% Mo, 0.185 wt% O, and 0.014 wt% N. The material was encapsulated in a fused quartz (Ar atmosphere), solution treated at 900 °C for 4 h and quenched in water. Cylindrical samples (10 mm × 15 mm) were subjected to three thermo-mechanical tests in the Gleeble® 3800 system pre-flushed with argon. All samples were heated at 50 °C/min up to 550 °C, held for 10 min and water quenched. The temperature was controlled by a K-thermocouple spot-welded directly to the sample. The first sample was heat-treated under minimal applied stress (0 MPa), the second one under elastic stress of 200 MPa. The third sample was heated to 550 °C and compressed to the true strain of 0.5 at the true strain rate of $10^{-3}$ s$^{-1}$ (referred to as Deformed). Deformation at elevated temperature lasted 10 min, matching the exposure time.

The microstructure was investigated by the Apreo 2 (Thermo Fisher Scientific) SEM operated at 2 keV to minimize the interaction volume and to allow the observation of clear and sharp fine particles. X-ray diffraction (XRD) was performed using the Rigaku Rapid II diffractometer in the reflection geometry using a curved 2D plate detector and a sealed Mo tube X-ray source with the spot size of 0.3 mm. XRD patterns were obtained by integrating the measured Debye-Scherrer rings. The microstructure investigation was complemented by Vickers microhardness measurement (500 g load, 10 s dwell time). 25 indents were made and evaluated by an automatic microhardness tester Qness Q10a. Samples were ground with SiC papers (down to #2000) and polished with 20 % $H_2O_2$ in OP-S colloidal silica.

## 3. Experimental results and discussion

The microstructure of the reference, solution treated sample (not presented here) consists of β grains with the size of 100 μm. It is known that material includes also nano-sized coherent particles of ω athermal ($\omega_{ath}$) phase [3]. BSE SEM micrographs of heat-treated samples are shown in Fig. 1. The microstructures of the sample annealed without external stress (0 MPa) (Fig. 1a) and the sample elastically loaded during annealing (200 MPa) (Fig. 1c) are similar. Low amount of thin α lamellae with a very high aspect ratio can be observed, mainly along/close to the β/β grain boundaries. Numerous fine particles (ellipsoids of the size in the range of tens of nm) uniformly dispersed in the β-matrix of both samples were identified by subsequent XRD measurements as the $\omega_{iso}$ phase, cf. Fig. 1b, d. On the other hand, completely different microstructure with a high fraction of tiny α phase particles was observed in the sample deformed in compression, appearing dark in the SEM image shown in Fig. 1f. The coarser band in the right-hand side of Fig. 1e corresponds to the original β/β grain boundary. Precipitation of α phase particles is enhanced at and around the boundary as shown previously also in ultra-fine grained Ti15Mo [15].

XRD patterns of all investigated states are shown in Fig. 2. The pattern of the ST sample shows narrow β matrix peaks and small and very broad peaks due to nanometer-sized $\omega_{ath}$ particles. After annealing without load (0 MPa), formation of the $\omega_{iso}$ phase is demonstrated by more pronounced ω phase peaks. The intensity of the $\omega_{iso}$ peaks further increased, though only slightly, after annealing with the applied elastic load of 200 MPa (see the cyan arrows in Fig. 2), indicating that the fraction of the $\omega_{iso}$ phase increases compared to the 0 MPa sample. On the other hand, no $\omega_{iso}$ phase was detected in the deformed sample. The microstructure of the deformed sample consists only of α and β phases (magenta arrows in Fig. 2).

Plastic deformation clearly promoted the β → α transformation in the deformed sample. The resulting microstructure is extremely fine with a large number of very small (less than 100 nm) α particles. This is caused most likely by dislocation activity which facilitates nucleation and accelerates the growth kinetics of α phase in the deformed condition/sample, as reported also in [7,10]. The morphology of α particles resembles nearly the globularized microstructure, which can be obtained in Ti17 alloy from initial α + β microstructure, using the same strain rate, but at much higher temperatures (e.g. 810 °C) [17].

The effect of the phase content on the mechanical properties of the metastable Ti-15Mo alloy was investigated by Vickers microhardness (HV) measurement. As it is apparent from the HV results plotted in

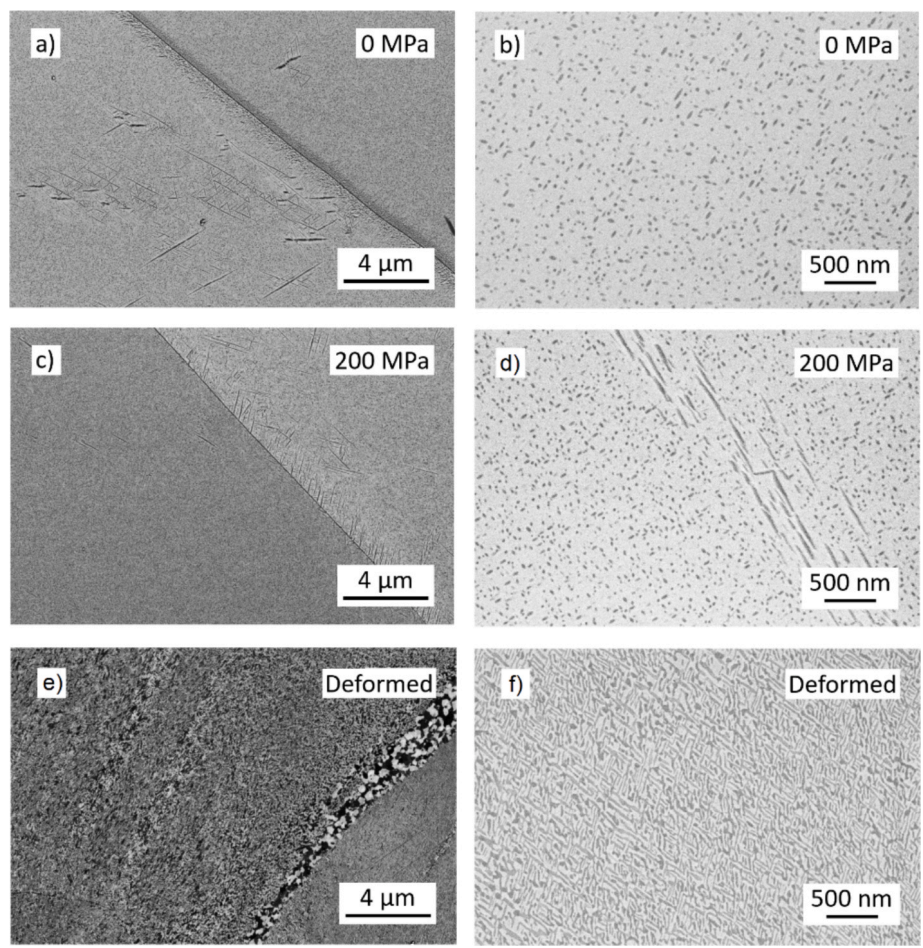

**Fig. 1.** A backscatter SEM micrographs of the samples annealed at 550 °C for 10 min (a,b) at 0 MPa, (c,d) under elastic load of 200 MPa and (e,f) deformed in compression (10 min). Loading direction is horizontal.

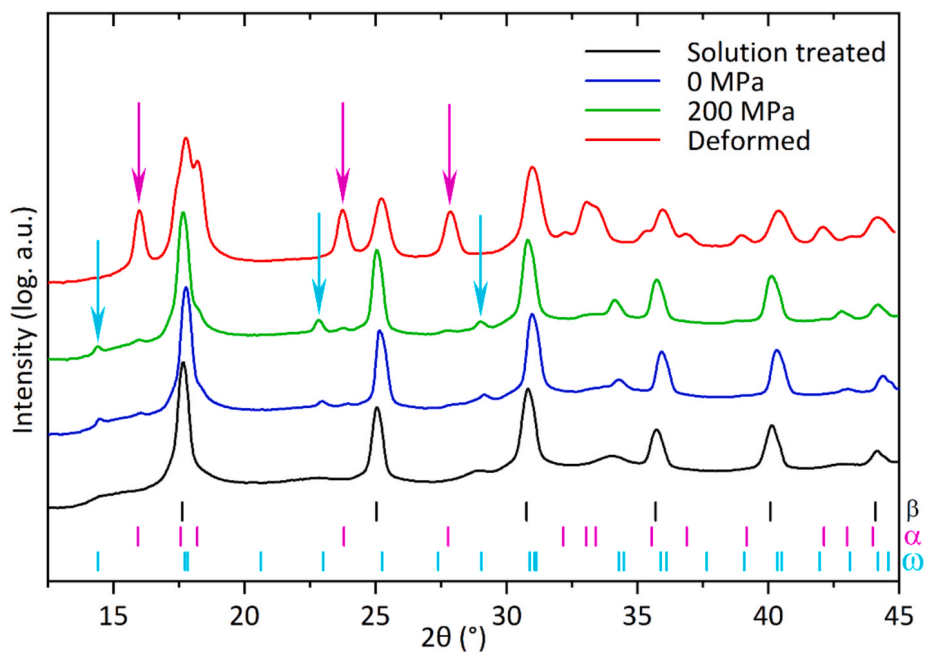

**Fig. 2.** Normalised XRD patterns of all conditions. The intensity is plotted in the logarithmic scale, and the patterns are shifted vertically for clarity. The positions of the peaks corresponding to α, β and ω phases are indicated in the lower part of the graph. Magenta and cyan arrows point to the most distinct peaks of α and ω phases, respectively. (For interpretation of the references to colour in this figure legend, the reader is referred to the web version of this article.)

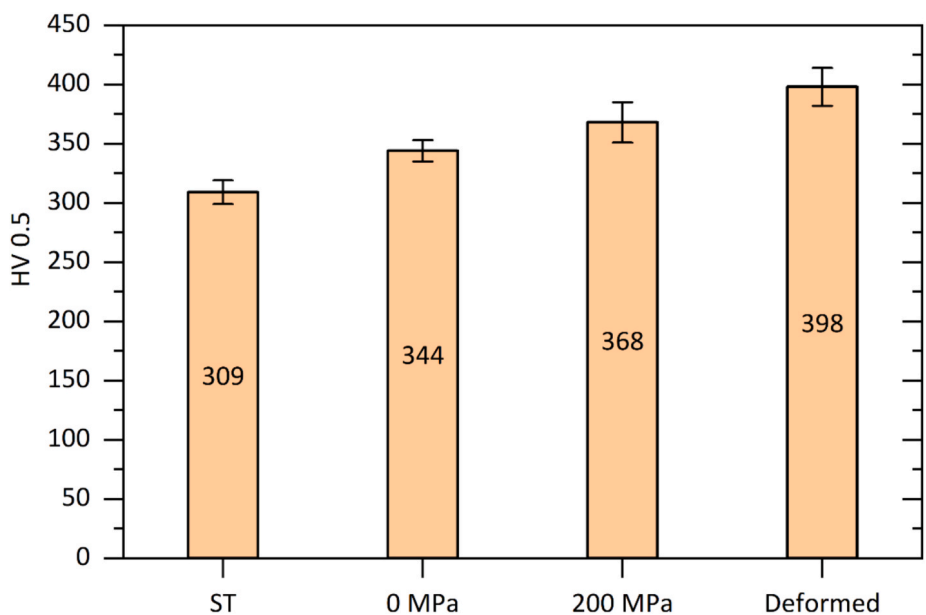

**Fig. 3.** Microhardness values of all investigated states.

Fig. 3, the microhardness of the ST sample increases from 309 ± 10 HV to 344 ± 9 HV upon annealing without stress (0 MPa). This can be attributed mainly to the $\omega_{iso}$ phase formation, while the effect of the formation of sparse α lamellae is minor. After annealing during elastic loading at 200 MPa, the microhardness increases to 368 ± 17 HV due to the more pronounced $\omega_{iso}$ phase formation. Based on image analysis of SEM images, the sample elastically loaded at 200 MPa exhibits smaller $\omega_{iso}$ particles, as compared to the sample annealed at 0 MPa, while both conditions contain comparable volume fraction of $\omega_{iso}$ of around 10 % (determined from XRD). It must be noted that minor differences in observed density, size and shape may not be significant as the appearance of $\omega_{iso}$ particles differs in each grain due to its orientation with respect to the sample surface. Nevertheless, microhardness measurements also suggest that $\omega_{iso}$ phase formation is more pronounced under the applied elastic stress. $\omega_{iso}$ phase formation upon (hydrostatic) pressure in (very) pure Ti was demonstrated decades ago [18], however at pressures by two orders of magnitude higher. More detailed analysis of this so far unresolved phenomenon is, however, beyond the scope of this Letter.

The deformed sample exhibits the highest microhardness of 398 ± 16 HV, due to the refined α + β microstructure [19]. Extremely fine α particles, present only in the deformed sample, result in high microhardness which was found for β + $\omega_{iso}$ phase content [20].

## 4. Conclusions

The influence of the mechanical loading on phase transformations in the solution treated Ti-15Mo alloy was investigated. The following conclusions can be drawn from this experimental study:

- Mechanical loading below yield stress of the material has a limited effect on phase transitions at 550 °C.
- Deformation applied during annealing promotes α precipitation.
- Ultra-fine microstructure of α particles smaller than 100 nm is formed.
- Microhardness increases from 309 HV (solution treated sample) to 398 HV (deformed sample).
- Careful temperature and deformation control may allow achieving unprecedented microstructural and mechanical conditions in metastable β-Ti alloys.

## CRediT authorship contribution statement

**Petr Harcuba:** Writing – review & editing, Investigation, Conceptualization. **Tomáš Krajňák:** Writing – original draft, Investigation. **Dalibor Preisler:** Investigation. **Jiří Kozlík:** Visualization, Data curation. **Josef Stráský:** Writing – review & editing. **Jana Šmilauerová:** Validation, Methodology. **Fernando Gustavo Warchomicka:** Writing – review & editing, Supervision. **Miloš Janeček:** Writing – review & editing, Project administration.

## Declaration of competing interest

The authors declare that they have no known competing financial interests or personal relationships that could have appeared to influence the work reported in this paper.

## Acknowledgements

Financial support by the Czech Science Foundation and the Austrian Science Fund under the joint project 22-21151K (CSF)/I 5818-N (FWF) is gratefully acknowledged. P. H., T. K., D. P., J. K., J. S., J. Š. and M. J. gratefully acknowledge the partial financial support by OP Johannes Amos Comenius of the MEYS of the CR, project No. CZ.02.01.01/00/22_008/0004591. The XRD measurements were performed in MGML (mgml.eu) within the program of Czech Research Infrastructures, project No. LM2018096.

## Data availability

The data that support the findings of this study will be openly available in the Zenodo repository at https://doi.org/10.5281/zenodo.14653251 under the CC-BY 4.0 license.